\documentclass[12pt,titlepage, authoryear]{article}

\usepackage{amssymb}

\usepackage{amsfonts,amsthm,graphicx,threeparttable,latexsym}
\usepackage[title]{appendix}
\RequirePackage[OT1]{fontenc}
\RequirePackage{graphicx}
\RequirePackage{amsthm}
\RequirePackage[cmex10]{amsmath}
\RequirePackage{natbib}
\RequirePackage[colorlinks,citecolor=blue,urlcolor=blue]{hyperref}
\RequirePackage{amssymb}
\theoremstyle{plain}

\newtheorem{thm}{Theorem}[section]

\newtheorem{lem}{Lemma}[section]
\newtheorem{definition}{Definition}[section]
\newtheorem{rmk}{Remark}[section]
\newtheorem{condition}{Condition}
\newtheorem{pf}{Proof}
\newtheorem{corollary}{Corollary}[section]

\providecommand{\abs}[1]{\lvert#1\rvert}
\providecommand{\norm}[1]{\lVert#1\rVert}
\begin{document}


\title{Nonlocal Solutions to  
 Dynamic Equilibrium Models: The Approximate Stable Manifolds Approach}
\author{Viktors Ajevskis\footnote{email: Viktors.Ajevskis@bank.lv}\\Bank of Latvia\footnote{The views  expressed in this paper are the sole responsibility of the author and do not
necessarily reflect the position of the Bank of Latvia.}}
\maketitle

\begin{abstract}
This study presents a method for constructing a sequence of approximate solutions of increasing accuracy to general equilibrium models on nonlocal domains. The method is based on a technique originated from dynamical systems theory. The approximate solutions are constructed employing the Contraction Mapping Theorem and the fact that solutions to general equilibrium models converge to a steady state. The approach allows deriving the a priori and a posteriori approximation errors of the solutions. Under certain nonlocal conditions we prove the convergence of the approximate solutions to the true solution and hence the Stable Manifold Theorem. We also show that the proposed approach can be treated as a rigorous proof of  convergence for the extended path algorithm to the true solution in a class of nonlinear rational expectation models.
\end{abstract}


%

%
%
%

\section{Introduction}
\label{intro}

The contribution of this study is as follows. First, based on the notion of stable manifold originated from dynamical systems theory  \citep{r30,r9} we propose  a method for constructing a sequence of approximate solutions of increasing accuracy to general equilibrium perfect foresight models on nonlocal domains. Second, we prove the convergence of the proposed approximate solutions to the true one, pointing out approximation errors and domain where solutions are defined. In this way, the result is a new proof of the Stable Manifold Theorem \citep{r31,r9}. Third, we establish  the relations between  the approach presented in this paper  and the extended path method (EP) proposed by  \cite{r15}, and show that the approach can be considered as a rigorous proof of convergence of  the EP algorithm. 


A stable manifold is a set of points that approach the saddle point as time tends to infinity \citep{r30,r9}. In fact, the set of solutions to a nonlinear rational expectations model determines the stable manifold because each solution must satisfy the stability condition, i.e. the convergence to the steady state in the long-run. In economic literature, this set is represented by a graph of a policy function (in other words, a decision function) that maps the state variables into the control variables. Since we primarily focus on constructing  approximations of  stable manifolds (not the manifolds themselves), we shall call the approach proposed in this paper the Approximate Stable Manifolds (ASM) method.

The existence of stable manifolds in the dynamical systems literature is proved using either the Lyapunov-Perron method or the graph transformation method of Hadamard \citep{r28,r44}. The Hadamard method constructs the entire stable manifold as the graph of a mapping, whereas the Lyapunov-Perron method relies on a dynamical characterization of stable manifolds as a set of initial points for solutions decaying exponentially to the steady state. 
The method developed in this paper has features of both the Hadamard and Lyapunov-Perron approaches. On the one hand, the iteration scheme of the ASM method can be treated as an implicit iteration scheme of the Hadamard method. On the other hand, the solution obtained by the ASM method is a fixed point of the so-called truncated Lyapunov-Perron operator \citep{r28}. We also show the relation between the Lyapunov-Perron approach and forward shooting method used for solving nonlinear rational expectation models \citep{r13} and relaying on this  explain the computational instability of the later. 

%
%


Initially, the proposed method involves the same steps as  perturbation methods \citep{r7,r3,r14}): (i) find a steady state; (ii) linearize the model around the steady state; (iii) decompose the Jacobian matrix at the steady state into stable and unstable blocks. The next step is to project the original system on the stable eigenspace (spanned on the stable eigenvectors) and the unstable one (spanned on the unstable eigenvectors). As a result, the system will be presented by two subsystems interrelated only through nonlinear terms. These terms are obtained as residuals after subtraction of the linearized system from the original nonlinear one; hence they vanish, together with their first derivatives, at the origin. Such a transformation makes the obtained system convenient to take the next stage of the method. Specifically, the approximate solutions are constructed by employing the convergence of solutions to the steady state and the Contraction Mapping Theorem. In this way we obtain a sequence of policy functions of increasing accuracy in a nonlocal domain.

The main results of the paper are Theorem~\ref{Thm1} and Theorem~\ref{Thm2}, which are proved in Section~\ref{theory}.  Theorem~\ref{Thm1} 
establishes the existence of a sequence of approximate policy functions.  Theorem~\ref{Thm2}  estimates the accuracy of the approximate solutions and the following corollaries prove the convergence of the approximate policy functions to the true one in a definite nonlocal domain. 
Thus, the result can be treated as a new proof of the Stable Manifold Theorem \citep{r30,r9}.
%


The proposed approach relates to the extended path method \citep{r15}. 
Specifically, under the assumption that the state variables are exogenous, at each point in time the solution of the extended path method applied to the transformed system lies on the corresponding ASM. In this way the EP method can be easily put in the ASM framework.
\cite{r27} mention that there is no proof that the EP algorithm converges to the true rational expectation solution.
The AMS method can be considered as a rigorous proof of convergence of the EP algorithm. 
Indeed, Theorems~\ref{Thm1} and~\ref{Thm2} 
can be applied directly for proving the convergence of the EP algorithm for the original (non-transformed) system. However, the use of linearization and spectral decomposition transforms the system into the form more convenient for verifying the conditions of the Contraction Mapping Theorem. Moreover, this gives us an advantage of the ASM approach over the EP in terms of the speed of convergence to the true solution. This occurs due to the fact  that the transformation makes the Lipschitz constant for the mappings involved smaller in a set of neighborhoods of the steady state.
Exploiting the Contraction Mapping Theorem also allows for obtaining the a priori and a posteriori approximation errors \citep{r22}. 
The analogy with the EP method  suggests that stochastic simulation in the ASM approach can be performed much as in \cite{r15}.

An attractive feature of the method is its capability to handle non-differentiable problems such as occasionally binding constraints, for example, the zero lower bound problem. 
This feature results from the fact that the Contraction Mapping Theorem requires a less restrictive condition than differentiability; namely,  this condition is  
the Lipschitz continuity.

The deterministic neoclassical growth model of \cite{r2}, considered as an example to illustrate how the method works, shows that  a first few approximations yield very high global accuracy. Even within the domain of convergence of the Taylor series  the low-order 
ASM solutions are more accurate than the high-order 
Taylor series approximations.

The rest of the paper is organized  as follows. The next section presents the model and its transformation  into a convenient form to deal with. In  Section~\ref{theory} our main results are stated and proved. Section~\ref{EP} establishes the relation between the proposed approach and  EP method. The ASM method is applied to neoclassical growth model in Section~\ref{example}.
Conclusions are presented in Section~\ref{conclusion}.

\section{The Model}
This paper is primarily concerned with the deterministic perfect foresight equilibrium of the models in the form:
\begin{eqnarray}
&f(y_{t+1},y_{t},x_{t+1},x_{t},z_{t})  =  0,\label{eq:3}\\
&z_{t+1}  =  \Lambda z_{t},\label{eq:4}
\end{eqnarray}
where  $x_t$ is an $n_x\times 1$ vector of endogenous  state variables at time $t$; $y_t$ is an $n_y\times 1$ vector containing $t$-period endogenous variables that are not state variables; $z_t$ is an $n_z\times 1$ vector of exogenous  state variables at time $t$;  $f$ maps ${\mathbb{R}}^{n_y}\times{\mathbb{R}}^{n_y}\times{\mathbb{R}}^{n_x}\times{\mathbb{R}}^{n_x}\times{\mathbb{R}}^{n_{z}}$   into ${\mathbb{R}}^{n_y}\times{\mathbb{R}}^{n_x}$ and is assumed to be at least twice 
continuously differentiable. 
All eigenvalues of matrix $\Lambda$ have modulus less than one. In the same way as in the EP method of \cite{r15}, the vector $z_t$ can be treated as initial values of temporary shocks at the time $t$.
We define the 
steady state as vectors $(\bar{y}, \bar{x}, 0)$  such that
\begin{equation}
    f(\bar{y},\bar{y}, \bar{x}, \bar{x}, 0)  =  0. \label{eq:7}
\end{equation}
Below, we also impose the conventional Blanchard-Kan condition \citep{r25} at the steady state. The problem then 
is to find a bounded solution $(x_{t}, y_{t})$ to \eqref{eq:3} 
for a given initial condition $(x_{0}, z_{0})$ for all $t \in \mathbb{N}$. 

\subsection{Model Transformation  and Preliminary Consideration }

By $(\hat{y_t}, \hat{x_t}, z_t)$ denote  vectors of deviation from the steady state. Linearizing (\ref{eq:3}) around the steady state, we get 
\begin{equation}
    f_1\hat{y}_{t+1} + f_2\hat{y_{t}} +f_3\hat{x}_{t+1} + f_4\hat{x_{t}} + f_5z_{t}+ N(\hat{y}_{t+1},\hat{y_{t}},\hat{x}_{t+1},\hat{x_{t}}, z_{t}) =  0, \label{eq:8}
\end{equation}
where $f_i$, $i = 1 \div 5$, are partial derivatives of the mapping $f$ with respect to $y_{t+1}$, $y_t$, $x_{t+1}$, $x_t$, and $z_t$, respectively, at $(\bar{y},\bar{y}, \bar{x}, \bar{x}, 0)$; and  $N$ is defined by 
\begin{equation*}
\begin{split}
   N(\hat{y}_{t+1},\hat{y_{t}},\hat{x}_{t+1},\hat{x_{t}},z_{t}) &= f(\bar{y}+\hat{y}_{t+1},\bar{y}+\hat{y_{t}},\bar{x}+\hat{x}_{t+1},\bar{x}+\hat{x_{t}},z_{t})\\
&-f_1\hat{y}_{t+1} - f_2\hat{y_{t}} - f_3\hat{x}_{t+1} - f_4\hat{x_{t}} - f_5\hat{z_{t}}.
\end{split}
\end{equation*}
The mapping $N$ is referred to as the nonlinear part of $f$.  By assumption on $f$, the mapping $N$ is continuously differentiable and vanishes, together with its first derivatives, at  $(0,0,0,0,0)$. For the sake of simplicity we assume that (\ref{eq:8})  can be transformed in such a way that $N$ does not depend on $\hat{y}_{t+1}$ and $\hat{x}_{t+1}$
\begin{equation}
f_1\hat{y}_{t+1} + f_2\hat{y_{t}} +f_3\hat{x}_{t+1} + f_4\hat{x_{t}} + f_5\hat{z_{t}}+ N(\hat{y_{t}},\hat{x_{t}}, z_{t}) =  0.\label{eq:9}
\end{equation}
This transformation can be done for many deterministic general equilibrium models (see Section~\ref{example} for the neoclassical growth model). 
Indeed, 
%
Equations  (\ref{eq:4})   and (\ref{eq:9}) can be written in the vector form as:
\begin{equation*}
   \Phi w_{t+1} =  \Gamma w_{t} + \begin{pmatrix} 0\\ N(w_{t})\end{pmatrix},
\end{equation*}
where 
\[w_{t}=(z_t, \hat{x_{t}}, \hat{y_{t}})',  \Phi=\left(
\begin{array}
[c]{ccc}%
I & 0 & 0\\
0 & f_{3} & f_{1}%
\end{array}
\right)\] 
and 
\[\Gamma=\left(
\begin{array}
[c]{ccc}%
\Lambda & 0 & 0\\
f_{5} & f_{4} & f_{2}%
\end{array}
\right).\]
We assume that the matrix $\Phi$ is invertible.\footnote{This assumption is made for ease of exposition. If $\Phi$ is a singular matrix, then in the sequel we must use a generalized eigenvalue decomposition as in \cite{r11}.} Then multiplying both sides
of the last equation by $\Phi^{-1}$ gives
\begin{equation}
    w_{t+1} = Kw_{t} + N_{1}(w_{t}),\label{eq:10}
\end{equation}
where 
%
 \[K = \Phi^{-1}\Gamma=\left(\begin{array}[c]{ccc}%
\Lambda&0&0\\
(f_{3},f_{1})^{-1}f_{5}&(f_{3},f_{1})^{-1}f_{4}&(f_{3},f_{1})^{-1}f_{2}%
\end{array}
\right)\]
and 
\[ N_{1}(w_{t})= \left(
\begin{array}
[c]{ccc}%
0\\(f_{3},f_{1})^{-1}N(w_{t})%
\end{array}
\right).\]

If the mapping $N$ does depend on $\hat{x}_{t+1}$ and $\hat{y}_{t+1}$, then we can employ the  Implicit Function Theorem to express  $\hat{x}_{t+1}$ and $\hat{y}_{t+1}$ as mappings of $\hat{x_{t}}, \hat{y_{t}}$ and $z_{t}$ 
Indeed, taking into account that the derivatives of $N(\cdot)$ with respect to  $\hat{x}_{t+1}$ and $\hat{y}_{t+1}$ are zero at the origin and the invertibility of $\Phi$, we can easily see that the conditions of the Implicit Function Theorem hold. 
In the case of the singular matrix $\Phi$, we do not have problems with zero-eigenvalues, because they correspond to identities that do not contain the terms with $\hat{x}_{t+1}$ and $\hat{y}_{t+1}$. Therefore, in what follows we assume that we have the representation of the original model in the form \eqref{eq:10}.

Next, the matrix $K$ is transformed into  a block-diagonal one which can be obtained, for example, by using the block-diagonal Schur factorization\footnote{The function bdschur of Matlab Control System Toolbox performs this factorization.}
\begin{equation}
    K = ZPZ^{-1},\label{eq:11}
\end{equation}
 where
\begin{equation} \label{25_}
P=\left[%
\begin{array}{cc}
{A} & {0} \\ 
{0} & {B}%
\end{array}%
\right]; 
\end{equation}
where $A$ and $B$ are quasi upper-triangular matrices with eigenvalues larger and smaller than one (in modulus), respectively;
and $Z$ is an invertible matrix.\footnote{The Jordan canonical form is also possible here, but it is not accurate computationally \citep{r26}.}
%
 We now introduce new variables 
\begin{equation}\label{eq:11_0}
(u_t, v_t)'=Z^{-1}(z_t,\hat{x_{t}},\hat{y_{t}})'
\end{equation}
%
Multiplying (\ref{eq:10}) by $Z^{-1}$ yields 
\begin{align}
u_{t+1}& = Au_{t} + F(u_{t}, v_{t})\label{eq:12a}\\
v_{t+1}& = Bv_{t} + G(u_{t}, v_{t})\label{eq:12b}
\end{align}
where $u_{t}\in{\mathbb{R}}^{n_{x}+n_{z}}$, $v_{t}\in{\mathbb{R}}^{n_{y}}$ and $( F(u_{t}, v_{t}),  G(u_{t}, v_{t}))^{'}=Z^{-1}N_{1}(Z\cdot(u_{t},v_{t})^{'})$. 
By construction, it follows that 
\begin{equation}
   F(0,0)=0,\quad G(0,0)=0,\quad  F^{'}(0,0)=0,\quad G^{'}(0,0)=0,\label{eq:13}
\end{equation}
where $F^{'}(0,0)$ and $G^{'}(0,0)$ stand for the Jacobian matrix of the mappings $F$ and $G$, respectively, at the point $(0,0)$.
The system in the form (\ref{eq:12a})--(\ref{eq:12b}) is convenient for obtaining the theoretical results of Section~\ref{theory}.

\section{Theoretical results}\label{theory}
The next subsection introduces some notation that will be necessary further on. 
Theorem~\ref{Thm1}  proves the existence of  a sequence of approximate policy functions in  Subsection~\ref{existence}, 
whereas Theorem~\ref{Thm2}   tells us about the accuracy of these approximations in  Subsection~\ref{accuracy}. 
\subsection{Notation and Definition}
By $U_{r_{u}}$ and $V_{r_{v}}$ denote the closed balls of radii $r_u$ and $r_v$ centered at the origin of ${\mathbb R}^{n_{x}+n_{z}}$ and ${\mathbb R}^{n_{y}}$, respectively. Let $X_{r_{u},r_{v}}= U_{r_{u}}+V_{r_{v}}$ be the direct sum of these balls. By $\lvert  \cdot  \rvert$ denote the Euclidean norm in ${\mathbb R}^{n}$. The induced norm for a real matrix $D$ is defined by
\begin{equation}
\lVert D\rVert =\sup_{\lvert s\rvert = 1}
 \lvert Ds \rvert.\label{eq:14}
\end{equation}
The matrix $Z$ in (\ref{eq:11})   can be chosen in such a way that 
\begin{equation}
\lVert A\rVert < \alpha + \gamma < 1  \text{ and }  \lVert B^{-1}\rVert < \beta + \gamma < 1,\label{eq:15}
\end{equation}
where $\alpha$ and $\beta$ are the largest eigenvalues of the matrices $A$ and $B^{-1}$ (in modulus), respectively, and $\gamma$  is arbitrarily small. 
This follows from the same
arguments as in  \citet[\S IV 9]{r6}, where it is done for the Jordan matrix
decomposition. 

Let $F:{\mathbb{R}}^{n_{x}+n_{z}}\times{\mathbb{R}}^{n_y} \to {\mathbb{R}}^{n_{x}+n_{z}}$, $G:{\mathbb{R}}^{n_{x}+n_{z}}\times{\mathbb{R}}^{n_y} \to {\mathbb{R}}^{n_{y}}$ and $h:{\mathbb{R}}^{n_{x}+n_{z}} \to {\mathbb{R}}^{n_y}$ are continuously differentiable maps.  Define the following  norms:
\begin{equation*}
\lVert G\rVert_{X_{r_{u},r_{v}}}=\sup_{(u,v)\in{X_{r_{u},r_{v}}}}
\lvert G(u,v)\rvert \text{,}\quad \lVert h\rVert_{U_{r_{u}}}=\sup_{u\in{U_{r_{u}}}}
\lvert h(u)\rvert;
\end{equation*}
\begin{equation*}
\begin{split}
&\lVert G^{'}\rVert_{X_{r_{u},r_{v}}}=\sup_{(u,v)\in{X_{r_{u},r_{v}}}}\lVert G^{'}(u,v)\rVert,\quad\lVert F^{'}\rVert_{X_{r_{u},r_{v}}}=\sup_{(u,v)\in{X_{r_{u},r_{v}}}}
\lVert F^{'}(u,v)\rVert,  \text{and}\\
  & \lVert h^{'}\rVert_{U_{r_{u}}}=\sup_{u\in{U_{r_{u}}}}
\lVert h^{'}(u)\rVert,
\end{split}
\end{equation*}
where $G^{'}(u, v)$,  $F^{'}(u, v)$ and $h'(u)$  are the Jacobian matrices of $G(u,v)$,  $F(u, v)$ and $h(u)$ at $(u, v)$ and $u$, respectively. 
%
%
\begin{definition}\label{def}
A mapping $s: X \rightarrow Y$ is called Lipschitz continuous if there exists a real constant $L \geq 0$ such that, for all $x_1$ and $x_2$ in $X$,
\begin{equation*}
\lvert s(x_2)-s(x_1)\rvert \leqslant L\lvert x_2-x_1\rvert,  
\end{equation*}
Any such $L$ is referred to as a Lipschitz constant for the mapping $s$.  
\end{definition}

\subsection{The existence of  a Sequence of Approximate Policy Functions}\label{existence}

\begin{thm}\label{Thm1}
Let $X_{r_{u},r_{v}}$ be a domain of definition for the  mappings $F$ and $G$ in (\ref{eq:12a}) and (\ref{eq:12b})  such that the following conditions holds:
%

\begin{condition}\label{Con1}
\begin{equation*}
\lVert G\rVert_{X_{r_{u},r_{v}}}<\frac{1-\lVert B^{-1}\rVert}{\lVert B^{-1}\rVert} r_{v};
\end{equation*}
\end{condition}
\begin{condition}\label{Con2}
\begin{equation*}
L<\frac{1}{4}\left( \frac{1}{\lVert B^{-1}\rVert } - \lVert A\rVert  \right);
\end{equation*}
where 
\begin{equation}
L=max(\lVert G'\rVert_{X_{r_{u},r_{v}}},\lVert F'\rVert_{X_{r_{u},r_{v}}}) \label{eq:20};
\end{equation}
\end{condition}
\begin{condition}\label{Con3}
If $(u_t, v_t)\in{X_{r_{u},r_{v}}}$, then $u_{t+1}(u_t, v_t)=Au_t+F(u_t,v_t)\in{U_{r_{u}}}$.
\end{condition}
Then, there exists a sequence of the mappings $h_i:U_{r_{u}}\to V_{r_{v}}$, $i \in \mathbb{N}$, satisfying  the  recurrent equations:
   \begin{equation}
h_{i}(u)=-B^{-1}G(u,h_{i}(u)) +B^{-1}h_{i-1}(Au+F(u,h_{i}(u)))\label{eq:21}
\end{equation}
with the initial condition $h_0 \equiv 0$. Moreover, the following inequalities for the norm of the mappings $h_i$ and their derivatives hold: 
\begin{equation}
\norm{h_i}_{U_{r_{u}}}\leqslant (1-\norm{ B^{-1}}^{i+1})\frac{\norm{ B^{-1}}\cdot\norm{G}_{X_{r_{u},r_{v}}}}{1-\norm{ B^{-1}}} 
\label{eq:22}
\end{equation}
\begin{equation}
\lVert h^{'}_{i}\rVert_{U_{r_{u}}}\leqslant \frac{1-\lVert B^{-1}\rVert L}{\lVert B^{-1}\rVert L}.\label{eq:23}
\end{equation}
\end{thm}

\begin{rmk}\label{Remark}
The neighborhood $X_{r_{u},r_{v}}$ that satisfies  {\em Conditions}~\ref{Con1}--\ref{Con3}  always exists locally, because the mappings $G(u,v)$ and $F(u,v)$ vanish, together with their first derivatives, at $(0,0)$. Nonetheless, {\em Conditions}~\ref{Con1}--\ref{Con3}  are not local by themselves.
\end{rmk}
\begin{rmk}\label{Remark2}
From {\em Condition}~\ref{Con1} and \eqref{eq:15} it follows that the right hand side of \eqref{eq:22} satisfies the inequality
\begin{equation}
(1-\norm{ B^{-1}}^{i+1})\frac{\norm{ B^{-1}}\cdot\norm{G}_{X_{r_{u},r_{v}}}}{1-\norm{ B^{-1}}} < r_{v}.
\end{equation}
\end{rmk}

\begin{pf}
The proof is by induction on $i$. More precisely, using the Contraction
Mapping Theorem, we derive by induction on $i\in\mathbb{N}$ the existence of $h_{i}$ satisfying (\ref{eq:21}). 
To satisfy the conditions of the Contraction
Mapping Theorem, we need the estimates (\ref{eq:22})--(\ref{eq:23}) for the mappings $h_i$ on
each stage of the induction. 

Suppose that $i=1$. Let $T_{1,u}$ be the
parameterized mapping of \ $V_{r_{v}}$ to ${\mathbb{R}}^{n_{v}}$ such that
 \begin{equation}
T_{1,u}(v)=-B^{-1}G(u,v)\quad \text{for each} \quad u\in U_{r_{u}}.\label{eq:24}
 \end{equation}
We claim that $T_{1,u}$ maps the closed ball  $V_{r_{v}}$ into itself and has
the Lipschitz constant less than one, and thus satisfies the conditions of the
Contraction Mapping Theorem  \citep{r22}.
Then there exists a fixed point $h_{1}$ of $T_{1,u}$ such that
 \begin{equation}
h_{1}(u)=-B^{-1}G(u,h_{1}(u)), \quad u\in U_{r_{u}}.\label{eq:25}
 \end{equation}
 Notice that the dependence of $h_{1}$ on $u$
determines the mapping of $U_{r_{u}}$to  ${\mathbb{R}}^{n_{v}}$. If in addition this mapping satisfies the inequalities (\ref{eq:22}) and (\ref{eq:23}), then the
induction hypothesis will be proved for $i=1$. Indeed, taking norm of both sides
(\ref{eq:24})  and using the norm property and {\em Condition}~\ref{Con1}, we have 
 \begin{equation}
\left\Vert T_{1,u}\right\Vert _{V_{r_{v}}}\leqslant\left\Vert
B^{-1}\right\Vert \cdot\left\Vert G\right\Vert _{X_{r_{u},r_{v}}}<r_{v}.\label{eq:26}
 \end{equation}
This means that $T_{1,u}$ maps the closed ball $V_{r_{v}}$ into itself.  Our
task now is to show that $T_{1,u}$ is a contraction. The Jacobian matrix of $T_{1,u}$ is
 \begin{equation}
T_{1,u}^{\prime}(v)=-B^{-1}G_{v}^{^{\prime}}(u,v),\label{eq:27}
\end{equation}
where $G_{v}^{^{\prime}}(u,v)$   is the Jacobian matrix of the mapping $G$ with respect to $v$ at the point $(u,v)$.
Taking the norm of both sides of (\ref{eq:27}) and using (\ref{eq:20}), the norm property and {\em Condition}~\ref{Con2}, we obtain
\begin{equation}
\left\Vert T_{1,u}^{\prime}(v)\right\Vert \leqslant\left\Vert B^{-1}\right\Vert L<1, \text{for all}\quad(u,v)\in X_{r_{u},r_{v}}. \label{eq:28}
\end{equation}
The norm $\left\Vert T_{1,u}^{\prime}(v)\right\Vert $ is an upper bound for the
Lipschitz constant of $T_{1,u}$ in the domain $V_{r_{v}}$ \citep{r22}. Since the mapping
$T_{1,u}$ has the Lipschitz constant less than one and maps the closed ball
$V_{r_{u}}$ into itself, we see that by the Contracting Mapping Theorem,
$T_{1,u}$  has a unique fixed point $h_{1}$  in $V_{r_{v}}$ for each $u\in U_{r_{u}}$. This
implies that the mapping $h_{1}$  defined by (\ref{eq:25}) exists. From (\ref{eq:26})  it follows
that 
$h_{1}$ satisfies  inequality (\ref{eq:22}). It remains to check that
the norm of the derivative of $h_{1}$ satisfies  inequality (\ref{eq:23}).
Differentiating (\ref{eq:25}) with respect to $u$, we obtain 
\[h_{1}^{^{\prime}}(u)=-B^{-1}G_{u}^{^{\prime}}(u,h_{1}
(u))-B^{-1}G_{v}^{^{\prime}}(u,h_{1}(u))h_{1}^{^{\prime}}(u).\]
Taking norms and applying the triangle inequality and using the norm property gives
\begin{equation}
{\norm{h^{'}_{1}(u)}} \leqslant \norm{B^{-1}}\cdot\norm{G^{'}_{u}(u,h_{1}(u))}+ \norm{B^{-1}}\cdot\norm{G^{'}_{v}(u,h_{1}(u))}\cdot{\norm{h{'}_{1}(u)}}\label{eq:29}
\end{equation}
for all $u\in U_{r_u}$. Rearranging terms in (\ref{eq:29}) and taking into account  the definitions of ${\norm{h^{'}_{1}}}_{U_{r_{u}}}$ and 
${\norm{G^{'}_{u}}}_ {X_{r_{u},r_{v}}}$, {\em Condition}~\ref{Con2} and (\ref{eq:20}), we get 
\begin{equation*}
{\norm{h^{'}_{1}}}_{U_{r_{u}}} \leqslant (1 -  \norm{B^{-1}}L)^{-1}\cdot\norm{B^{-1}}L.
\end{equation*}
From {\em Condition}~\ref{Con2} it follows easily that $ \norm{B^{-1}}L < 1/2$. This implies that
\begin{equation*}
{\norm{h^{'}_{1}}}_{U_{r_{u}}} < (1 -  \norm{B^{-1}}L)/(\norm{B^{-1}}L),
\end{equation*}
hence $h_1$  satisfies the inequality (\ref{eq:24}). Therefore, the inductive assumption is proved for $i=1$.

Next, suppose inductively that there exist mappings $h_{k}(u), k =1,2,\dotsc, i$, that satisfy (\ref{eq:21})--(\ref{eq:23}). Let $T_{i+1,u}$ be the
parameterized mapping of $V_{r_{v}}$ to ${\mathbb{R}}^{n_{v}}$ such that
 \begin{equation}
T_{i+1,u}(v)=-B^{-1}G(u,v) +B^{-1}h_{i}(Au + F (u, v)) \label{eq:30}
 \end{equation}
for each $u\in U_{r_{u}}$.
As before, we shall show that $T_{i+1,u}$  satisfies the Contraction Mapping Theorem conditions. Indeed, taking norms in (\ref{eq:30}), using the norm property and applying the triangle inequality yields
 \begin{equation}
\norm{T_{i+1,u}}_{V_{r_{v}}}\leqslant\left\Vert
B^{-1}\right\Vert \cdot\left\Vert G\right\Vert _{X_{r_{u},r_{v}}} + \norm{B^{-1}}\cdot{\norm{h_{i}}}_{U_{r_{u}}}.\label{eq:31}
 \end{equation}
By the inductive assumption, inequality (\ref{eq:22}) holds; therefore 
  \begin{equation}
\begin{split}
\norm{ T_{i+1,u}} _{V_{r_{v}}}&\leqslant 
(1-{\norm{ B^{-1}}^{i+2}})\frac{\norm{ B^{-1}}\cdot\norm{G}_{X_{r_{u},r_{v}}}}{1-\norm{ B^{-1}}} <r_{v} ,\label{eq:32}
\end{split}
 \end{equation}
where the last inequality follows from Remark~\ref{Remark2}.
This means that 
\[T_{i+1,u}:V_{r_{v}}\rightarrow V_{r_{v}}\quad\text{for all}\quad  u\in U_{r_{u}}.\] 
The Jacobian matrix of the mapping $T_{i+1,u}$ at the point $v$ is
 \begin{equation}
T_{i+1,u}^{\prime}(v)=-B^{-1}G_{v}^{^{\prime}}(u,v) + B^{-1}h^{'}_{i}(Au+F(u,v))F^{'}_{v}(u,v) .\label{eq:33}
\end{equation}
Taking norms, using the norm property and applying the triangle inequality, {\em Condition}~\ref{Con3} and (\ref{eq:20}), we obtain 
 \begin{equation}
\norm{T_{i+1,u}^{\prime}(v)}\leqslant \norm{B^{-1}}L + \norm{B^{-1}}\cdot{\norm{h^{'}_{i}}}_{ U_{r_{u}}}L.\label{eq:34}
\end{equation}
Inserting the inductive assumption for ${\norm{h^{'}_{i}}}_{ U_{r_{u}}}$, we have $\norm{T_{i+1,u}^{\prime}(v)}<1$ for all $v\in V_{r_v}$. Since the mapping $T_{i+1,u}$ has the Lipschitz constant less than one and maps $V_{r_v}$ into itself, it has a unique fixed point $ h_{i+1}(u)$  for each $u\in U_{r_{u}}$. This implies that there exists a mapping $h_{i+1}$ of $U_{r_{u}}$ to $V_{r_{v}}$ such that 
 \begin{equation}
h_{i+1}(u)=-B^{-1}G(u,h_{i+1}(u)) +B^{-1}h_{i}(Au+F(u,h_{i+1}(u))).\label{eq:35}
\end{equation}
From (\ref{eq:32}) it follows that the norm of $h_{i+1}$ satisfies  inequity (\ref{eq:22}). To conclude the inductive assumption for $i+1$, it remains to check inequality (\ref{eq:23}) for the norm of the derivative of the mapping $h_{i+1}$. This is the harder part of the proof. Indeed, taking the  derivative of $h_{i+1}$ at the point $u$, we have
 \begin{equation*}
\begin{split}
&h_{i+1}^{\prime}(u)=-B^{-1}G_{u}^{\prime}(u,,h_{i+1}(u)) - B^{-1}G_{v}^{\prime}(u,h_{i+1}(u))h_{i+1,u}^{\prime}(u) \\
&+B^{-1}h_{i}^{\prime}(Au+F(u,h_{i+1}(u)))\left[A+F^{\prime}_u(u,h_{i+1}(u)) +F^{\prime}_v(u,h_{i+1}(u))h_{i+1,u}^{\prime}(u)\right].\label{eq:36}
\end{split}
\end{equation*}
Taking norms and using the triangle inequality and the norm property, we obtain
 \begin{equation*}
\begin{split}
\norm{h_{i+1}^{\prime}(u)}&\leqslant\norm{B^{-1}}\cdot\norm{G_{u}^{\prime}(u,,h_{i+1}(u))} + \norm{B^{-1}}\cdot\norm{G_{v}^{\prime}(u,h_{i+1}(u))}\cdot\norm{h_{i+1}^{\prime}(u)} \\
&+\norm{B^{-1}}\cdot\norm{h_{i}^{\prime}(Au+F(u,h_{i+1}(u)))}\cdot\norm{A+F^{\prime}_u(u,h_{i+1}(u))} \\
&+\norm{B^{-1}}\cdot\norm{h_{i}^{\prime}(Au+F(u,h_{i+1}(u)))}\cdot\norm{F^{\prime}_v(u,h_{i+1}(u))}\cdot\norm{h_{i+1}^{\prime}(u)}\label{eq:37}
\end{split}
\end{equation*}
for all $u\in U_{r_{u}}$. Using (\ref{eq:20}) and {\em Condition}~\ref{Con3} yields
\begin{equation*}
\begin{split}
\norm{h^{'}_{i+1}}_{U_{r_{u}}} &\leqslant \norm{B^{-1}}L + \norm{B^{-1}}\cdot L \cdot{\norm{h^{'}_{i+1}}}_{U_{r_{u}}} \\
&+ \norm{B^{-1}}\cdot{\norm{h^{'}_{i}}}_{U_{r_{u}}} \cdot (\norm{A}+L)+\norm{B^{-1}}\cdot{\norm{h^{'}_{i}}}_{U_{r_{u}}} \cdot L \cdot{\norm{h^{'}_{i+1}}}_{U_{r_{u}}}.\label{eq:38}
\end{split}
\end{equation*}
By the inductive assumption, the norm ${\norm{h^{'}_{i}}}_{U_{r_{u}}}$ satisfies the estimate (\ref{eq:23}), hence $(1-\norm{B^{-1}}L -\norm{B^{-1}}L \cdot{\norm{h^{'}_{i}}}_{U_{r_{u}}} )>0$. Then from (\ref{eq:23}) it follows that 
\begin{equation}
\norm{h^{'}_{i+1}}_{U_{r_{u}}} \leqslant \frac{ \norm{B^{-1}}L + \norm{B^{-1}}\cdot{\norm{h^{'}_{i}}}_{U_{r_{u}}} \cdot (\norm{A}+L)}{1-\norm{B^{-1}}L -\norm{B^{-1}} \cdot L \cdot{\norm{h^{'}_{i}}}_{U_{r_{u}}}}.\label{eq:39}
\end{equation}
Consider now the following difference equation:
\begin{equation}
s_{i+1}= \frac{ \rho +(\norm{B^{-1}}\cdot \norm{A}+\rho)s_{i}}{1-\rho-\rho s_{i}},\label{eq:40}
\end{equation}
where $\rho=\norm{B^{-1}}L.$ 
\begin{lem}
Suppose 
\begin{equation}
\rho < (1-\norm{B^{-1}}\cdot \norm{A})/4;\label{ineq:41}
\end{equation}
then the difference equation  (\ref{eq:40}) has two fixed points:
\begin{equation}
s^{*}_{1}= \frac{1-2 \rho - \norm{B^{-1}}\cdot \norm{A}- \sqrt{(1-2\rho-\norm{B^{-1}}\cdot \norm{A})-4\rho^2}}{2\rho} \label{eq:41}
\end{equation}
and
\begin{equation}
s^{*}_{2}= \frac{1-2 \rho - \norm{B^{-1}}\cdot \norm{A}+\sqrt{(1-2\rho-\norm{B^{-1}}\cdot \norm{A})-4\rho^2}}{2\rho}       \label{eq:42}
\end{equation}
such that 
\begin{equation}
s^{*}_{1} \leqslant s^{*}_{2}< \frac{1 - \rho}{\rho},  \label{eq:43}
\end{equation}
where $s^{*}_{1} $  is a stable fixed point, 
 $s^{*}_{2}$ is an unstable fixed point.  If  $s_0 = 0$, then $s_i, i = 1,2, \dots$, is a monotonically increasing sequence that converges to  $s^{*}_{1}$. 
\end{lem}
\begin{pf}
The lemma  can be proved by direct calculation. 
\end{pf}

Inequality (\ref{ineq:41}) easily follows from  {\em Condition}~\ref{Con2} of the theorem. Comparing (\ref{eq:39}) and (\ref{eq:40}) for the initial point $s_0 = 0$ and the initial mapping $h_0\equiv 0$, we have $s_i \geq \norm{h^{'}_{i+1}}_{U_{r_{u}}}$ for $i\in \mathbb{N}$, i.e. $\norm{h^{'}_{i+1}}_{U_{r_{u}}}$ is majorized by $s_i$.  From (\ref{eq:43}) it follows that 
\begin{equation*}
\norm{h^{'}_{i+1}}_{U_{r_{u}}} \leqslant s^{*}_{1} \leqslant \frac{1 - \rho}{\rho} = \frac{1-\norm{B^{-1}}L}{\norm{B^{-1}}L}.
\end{equation*}
Therefore, the mapping $h_{i+1}$ satisfies  inequality (\ref{eq:23}). This concludes the induction argument. \qed
\end{pf}


\subsection{The Accuracy of the Approximate Policy Functions and Their Convergence}\label{accuracy}
The condition for the graph of a mapping $h$ to be an invariant manifold is that the image under transformation (\ref{eq:12a})--(\ref{eq:12b}) of a general point of the graph of $h$ must again be in the graph of $h$. This holds if and only if
\begin{equation*}
Bh(u)+G(u,h(u)) = h(Au+F(u,h(u))).
\end{equation*}
Taking into account the invertibility of $B$, we have
\begin{equation}
h(u)=-B^{-1}G(u,h(u)) +B^{-1}h(Au+F(u,h(u))).\label{eq:18}
\end{equation}
Thus, the true policy function $h$ satisfies \eqref{eq:18}.
 The next theorem gives the estimate of the error created by the approximate policy functions $h_i$, $i\in \mathbb{N}$, obtained in Theorem~\ref{Thm1}.
\begin{thm}\label{Thm2}
Under the conditions of Theorem~\ref{Thm1},  
the following inequality holds: 
\begin{equation}
\abs{h_{n}(u_{t}) - h(u_{t})}\leqslant a^{n-1}\cdot\norm{B^{-1}}\cdot\left(1-\norm{B^{-1}}L\right)^{-1}\cdot\abs{h(u_{t+n})}, \label{eq:44}
\end{equation}
for all $u_t\in U_{{r}_u}$ and $n\in \mathbb{N}$, where 
\begin{equation}
a=\frac{2\norm{B^{-1}}}{1+\norm{B^{-1}}\cdot\norm{A}}, \label{a}
\end{equation}
and $u_{t+n+1}$ is the solution of the difference equation
\begin{equation}
 u_{s+1} = Au_{s} + F(u_{s}, h(u_{s}))\label{eq:45} 
\end{equation}
 at the time $t+n$ and with the initial value $u_{t}$.\footnote{In other words, $u_{t+n}$ is the $u$-coordinate of the solution lying on the stable manifold.}
\end{thm}
\begin{pf}
The proof is by induction on $i$ for $1\leqslant i \leqslant n$. Suppose that $i=1$. 
Let $u_{t+n-1}$ be the solution to (\ref{eq:45}) at the time $t+n-1$ for an initial point $u_t\in U_{r_{u}}$.
Subtracting (\ref{eq:18}) from (\ref{eq:25}) at the point $u_{t+n-1}$, taking norms, using the norm property and applying the triangle inequality, we get
\begin{equation}
\begin{split}
&\abs{h_{1}(u_{t+n-1}) - h(u_{t+n-1})}\leqslant\\
& \norm{B^{-1}}\cdot \abs{G(u_{t+n-1},h_{1}(u_{t+n-1})) - G(u_{t+n-1},h(u_{t+n-1}))}\\
 &+ \norm{B^{-1}}\cdot\abs{h( Au_{t+n-1} + F(u_{t+n-1}, h(u_{t+n-1})))}. \label{eq:46}
\end{split}
\end{equation}
It follows easily that 
\begin{equation}
\abs{G(u,h_{1}(u)) - G(u,h(u))}\leqslant L\cdot \abs{h_{1}(u) - h(u)}. \label{eq:47}
\end{equation}
Combining (\ref{eq:45}), (\ref{eq:46}) and (\ref{eq:47}), and taking into account the inequality $\norm{B^{-1}}L<1$  , we obtain
\begin{equation}
\abs{h_{1}(u_{t+n-1}) - h(u_{t+n-1})}\leqslant \left(1- \norm{B^{-1}}L\right)^{-1}\cdot\norm{B^{-1}}\cdot\abs{h( u_{t+n})}. \label{eq:48}
\end{equation}
Therefore, the inductive assumption is proved for $i=1$. 
Assume now that inequality (\ref{eq:44}) holds for $k = 1, 2,\dotsc, i$; we will prove it for $i+1$.
In particular, for $k= i$ we have 
\begin{equation}
\abs{h_{i}(u_{t+n-i}) - h(u_{t+n-i})}\leqslant a^{i
-1}\cdot\norm{B^{-1}}\cdot\left(1-\norm{B^{-1}}L\right)^{-1}\cdot\abs{h(u_{t+n})}. \label{eq:49}
\end{equation}
Subtracting (\ref{eq:18}) from (\ref{eq:21}) both written for the argument $u_{t+n-i-1}$ and denoting for brevity $j=n-i-1$, we get
 \begin{equation}
\begin{split}
&h_{i+1}(u_{t+j}) - h(u_{t+j})=-B^{-1}\big[G(u_{t+j},h_{i+1}(u_{t+j})) - G(u_{t+j},h(u_{t+j}))\\
&+h_{i}( Au_{t+j} + F(u_{t+j}, h_{i+1}(u_{t+j})))-h( Au_{t+j} + F(u_{t+j}, h(u_{t+j})))\big]. \label{eq:50}
\end{split}
\end{equation}
Adding and subtracting the term $h_{i}( Au_{t+j} + F(u_{t+j}, h(u_{t+j})))$ in brackets, we have
 \begin{equation}
\begin{split}
&h_{i+1}(u_{t+j}) - h(u_{t+j})=-B^{-1}\big[G(u_{t+j},h_{i+1}(u_{t+j})) - G(u_{t+j},h(u_{t+j}))\\
&+h_{i}( Au_{t+j} + F(u_{t+j}, h_{i+1}(u_{t+j})))-h_{i}( Au_{t+j} + F(u_{t+j}, h(u_{t+j})))\\
&+h_{i}( Au_{t+j} + F(u_{t+j}, h(u_{t+j})))-h( Au_{t+j} + F(u_{t+j}, h(u_{t+j})))\big]. \label{eq:51}
\end{split}
\end{equation}
Using (\ref{eq:20}), (\ref{eq:47})  and the triangle inequality gives 
 \begin{equation}
\begin{split}
&\abs{h_{i+1}(u_{t+j}) - h(u_{t+j})}\leqslant \norm{B^{-1}}\big[ L \abs{h_{i+1}(u_{t+j}) - h(u_{t+j})}\\
&+ L{\norm{h_{i}^{'}}}_{U_{r_{u}}}\cdot \abs{h_{i+1}(u_{t+j}) - h(u_{t+j})}\\
&+ \abs{h_{i}( Au_{t+j} + F(u_{t+j}, h(u_{t+j})))-h( Au_{t+j} + F(u_{t+j}, h(u_{t+j
})))}\big].\label{eq:53}
\end{split}
\end{equation}
Rearranging terms, we obtain
 \begin{equation}
\begin{split}
&\left(1-\norm{B^{-1}} L- \norm{B^{-1}}\cdot L\cdot{\norm{h_{i}^{'}}}_{U_{r_{u}}}\right)\cdot\abs{h_{i+1}(u_{t+j}) - h(u_{t+j})}\\
&\leqslant \norm{B^{-1}}\cdot\abs{h_{i}( u_{t+j+1}) -h( u_{t+j+1})}.\label{eq:53_}
\end{split}
\end{equation}
With the notation
\begin{equation}
b=1-\norm{B^{-1}} L- \norm{B^{-1}}\cdot L\cdot{\norm{h_{i}^{'}}}_{U_{r_{u}}},\label{eq:53_0}
\end{equation}
from  (\ref{eq:23}) we have $b>0$.
%
From \eqref{eq:53_} it follows that
 \begin{equation*}
\abs{h_{i+1}(u_{t+j}) - h(u_{t+j})}\ < b^{-1}\cdot\norm{B^{-1}}\cdot\abs{h_{i}( u_{t+j+1}) -h( u_{t+j})}.
\end{equation*}
Inserting the inductive assumption (\ref{eq:49}) for the upper bound of 
$\abs{h_{i}(u_{t+j+1})-h(u_{t+j+1})}$ yields
 \begin{equation}
\abs{h_{i+1}(u_{t+j}) - h(u_{t+j})} < b^{-1}a^{i
-1}{\norm{B^{-1}}}^{2}\left(1-\norm{B^{-1}}L\right)^{-1}\abs{h(u_{t+n})}. \label{eq:54}
\end{equation}
From the proof of Theorem~\ref{Thm1} it follows that ${\norm{h_{i}^{'}}}_{U_{r_{u}}}\leqslant s_{1}^{*}$, where $s_{1}^{*}$  is given by (\ref{eq:41}). Therefore
 \begin{equation}
b^{-1}\leqslant\left(1-\norm{B^{-1}} L- \norm{B^{-1}}\cdot Ls_{1}^{*}\right)^{-1}.\label{eq:55}
\end{equation}
by \eqref{eq:53_0}.
Consider now the following function of $L$:
\begin{equation*}
q(L)=\left(1-\norm{B^{-1}} L- \norm{B^{-1}}\cdot Ls_{1}^{*}\right)^{-1}.
\end{equation*}
Inserting $s_{1}^{*}$ from (\ref{eq:41}) gives
\begin{equation*}
q(L)=2\left[1+\norm{B^{-1}}\cdot \norm{A}+ \sqrt{\left(1-2\norm{B^{-1}}L-\norm{B^{-1}}\cdot \norm{A}\right)^{2}-4\left(\norm{B^{-1}}L\right)^{2}}\right]^{-1}.
\end{equation*}
It can easily be checked that 
the function $q(L)$  attains its maximum  in the interval $\left[0, \left(1- \norm{B^{-1}}\cdot \norm{A}\right)/4\right]$  at the point $L^{*} = \left(1- \norm{B^{-1}}\cdot \norm{A}\right)/4$,
and the value of $q$ at this point is 
\begin{equation*}
q^{*}=q\left(L^{*}\right)=\frac{2}{1+\norm{B^{-1}}\cdot \norm{A}}.
\end{equation*}
Then from (\ref{eq:55}) it follows that
 \begin{equation}
b^{-1}\leqslant \frac{2}{1+\norm{B^{-1}}\cdot \norm{A}}.\label{eq:56}
\end{equation}
Using (\ref{a}), (\ref{eq:54}) and (\ref{eq:56}) and returning the  index $i$, we obtain
 \begin{equation*}
\abs{h_{i+1}(u_{t+n-i-1}) - h(u_{t+n-i-1})} < a^{i}{\norm{B^{-1}}}\left(1-\norm{B^{-1}}L\right)^{-1}\abs{h\left(u_{t+n}\right)}. \label{eq:57}
\end{equation*}
This concludes the inductive steps. Finally, substituting $n$ for $i+1$ in the last equation
, we have (\ref{eq:44}). \qed
\end{pf}
\begin{corollary}\label{Corollary1}
As $n\to\infty$ the estimate (\ref{eq:44}) takes the form:
\begin{equation}
\abs{h_{n}(u_{t}) - h(u_{t})} <C_1 \left[a(\norm{A}+\theta)^2\right]^{n},
\label{eq:58}
\end{equation}
 for all $u_t \in U_{r_{u}}$, where $ C_1>0$ is some constant and $\theta> 0$ is an arbitrarily small constant.
\end{corollary}
\begin{pf}
From \citet[Corollary 5.1]{r6}  it follows that  for the solutions lying on the stable manifold the following estimate holds: 
\begin{equation}
\abs{u_{t+n}} <C_2 \left(\norm{A}+\theta \right)^{n},\label{eq:59}
\end{equation}
 where $ C_2$ is some constant and $\theta> 0$ is an arbitrarily small constant. 
Therefore, $\norm{A}+\theta <1$, and $u_{t+n}\to 0$ as $n\to\infty$. Since $h(0)=0$ and $h^{'}(0)=0$ (\citet[Lemma 5.1]{r6}), the Taylor expansion terms of the mapping $h$ start at least at the second order. From this and \eqref{eq:59} for $n$ sufficiently large we have
\begin{equation}
\abs{h(u_{t+n})} <C_3 (\norm{A}+\theta)^{2n},\label{eq:60}
\end{equation}
where $C_3$ is some constant. 
Combining the last inequality and (\ref{eq:44}), we obtain (\ref{eq:58}), where $C_1=C_3 a^{-1}\norm{B^{-1}}\left(1-\norm{B^{-1}}L\right)^{-1} $.\qed
\end{pf}

\begin{corollary}\label{Coll2}
Under the conditions of Theorem 1 the mappings $h_n$ tend to $h$ in the $C^0$-topology as  $n\to\infty$. 
\end{corollary}
\begin{pf}
It is easily shown that if $\norm{B^{-1}}<1$, $\norm{A}+\theta<1$,  and $\theta$ is sufficiently small, then taking into account \eqref{a}, we have
\begin{equation}
a(\norm{A}+\theta)^2=\frac{2\norm{B^{-1}}\cdot (\norm{A}+\theta)^2}{1+\norm{B^{-1}}\cdot\norm{A}}<1,\label{eq:61}
\end{equation}
The assertion of the corollary now follows from (\ref{eq:58}). \qed
\end{pf}
\begin{rmk}\label{rem3.2}
 {\em 
From \eqref{eq:23} it follows that  $h$ is a Lipschitz-continuous function.
}
\end{rmk}
\begin{rmk}
 {\em 
The Stable Manifold Theorem  \citep{r30,r31} follows from Corollary~\ref{Coll2} and Remark~\ref{rem3.2}.
}.
\end{rmk}
\begin{rmk}
{\em 
The right-hand side of (\ref{eq:18}) can be taken as defining a nonlinear operator on functions $h$, i.e.  \eqref{eq:18} can be interpreted as the fixed-point equation $h = Th$, where
\begin{equation*}
Th(u)=-B^{-1}G(u,h(u)) +B^{-1}h(Au+F(u,h(u))).
\end{equation*}
The approach related to finding the fixed point of $T$ to construct the stable manifold is called the Hadamard method in the dynamical systems literature \citep{r28,r44}. The idea is to choose a functional space on which the operator $T$ is a contraction. 
Then, its iterations give explicitly approximate solutions
\begin{equation}
h_{i}(u)=-B^{-1}G(u,h_{i-1}(u)) +B^{-1}h_{i-1}(Au+F(u,h_{i-1}(u))).\label{eq:18_}
\end{equation}
Comparing  \eqref{eq:18_} and \eqref{eq:21}, one can treat the proposed recurrent procedure \eqref{eq:21} as an implicit iteration scheme of the Hadamard method. 
} 
\end{rmk}
\begin{rmk}
{\em 
It is easily shown that iterating \eqref{eq:21} one can obtain the mapping $h_i$, $i=1,2,\dots,$ as a fixed point of the parameterized mapping 
\begin{equation}\label{LP}
T_{i,v}^{LP}(u_{t})=-\sum_{k=0}^{i}B^{-k-1}G(u_{t+k}(u_t,v_t),v_{t+k}(u_t,v_t)),
\end{equation}
where $(u_{t+k}(u_t,v_t),v_{t+k}(u_t,v_t))$ is the solution of the system \eqref{eq:12a}--\eqref{eq:12b} with the initial condition $(u_t,v_t)$. 
In the dynamical systems literature, the approach based on seeking a fixed point of the operator $T_{\infty,v}^{LP}$ is called the Lyapunov-Perron method and the operator $T_{i,v}^{LP}$ is called the truncated Lyapunov-Perron operator \citep{r28}. 
In fact,  the Lyapunov-Perron method is the shooting method applied to the transformed system \eqref{eq:12a}--\eqref{eq:12b}.\footnote{For application to non-transformed models see, for example, \cite{r13}.}
The major shortcoming of this approach is computational instability because the coordinate $u_{t+k}$ of a solution  grows exponentially fast 
with $k$ if an initial point is not on the stable manifold.  This requires  great care about choosing a starting point of iterations.
} 
\end{rmk}

\subsection{Initial Conditions for the Transformed System and Recovery of the Original Variables}\label{init}
To obtain 
a solution to the system \eqref{eq:12a}--\eqref{eq:12b} we need to find the initial condition $u_0$  corresponding to the initial condition $(x_0,z_0)$ for the original system \eqref{eq:3}--\eqref{eq:4}.
Using the transformation $Z$ from \eqref{eq:11} and \eqref{eq:11_0} at the time $t=0$, gives the system of equations
\begin{align}
Z_{11}u_0+Z_{12}h_{i}(u_0)&=z_0,\label{init1}\\ 
Z_{21}u_0+Z_{22}h_{i}(u_0)&=x_0-\bar{x}. \label{init2}
\end{align}
The initial condition $u_0$ then can be found by solving (\ref{init1})--(\ref{init2}). It is not hard to prove  that under the conditions of  Theorem~\ref{Thm1} the system of equations \eqref{init1}--\eqref{init2} has a unique solution $u_0$. 
Having the mapping $h_i$ and knowing the initial condition $u_0$, we can find the approximate equilibrium trajectory  from the system:
\begin{align}
u_{t+1}& = Au_{t} + F(u_{t},h_{i}( u_{t})),\label{eq:12a_}\\
v_{t+1}& =h_{i}( Au_{t} + F(u_{t},h_{i}( u_{t}))),\label{eq:12b_}
\end{align}
for $t=0,1,\ldots$
Using the transformation $Z$
, we  recover the dynamics of the original variables 
\begin{equation}\label{transf}
\begin{pmatrix}z_t\\x_t\\y_t\end{pmatrix}=
Z\begin{pmatrix}{u}_{t}\\h_i(u_t)\end{pmatrix}+
\begin{pmatrix}0\\\bar{x}\\\bar{y}\end{pmatrix}.
\end{equation}

\section{Connection with the EP Method. Stochastic simulation}\label{EP}
The proposed approach bears a similarity to that of the EP method \citep{r15}. 
%
Indeed, for simplicity we  assume that the variable $x_t$ is exogenous, then it is easily shown that  the mapping $F$ in \eqref{eq:12a} has the form $F(u, v)=F(u)$. 
Therefore, $u$ is also an exogenous variable, and thus
 under an initial condition $u_t$ one can find the solution $u_{t+i}$ for all $i \in \mathbb{N}$. Under this assumption the EP methods  applied to the transformed system  (\ref{eq:12a})--(\ref{eq:12b}) involves the following steps:

\begin{enumerate}
\item Fix a horizon $n$, and the terminal value $v_{t+n+1} = 0$.
\item Make a guess $V^{0}_{n, t+i}=0$ for $v_{t+i}, i=1,\ldots n.$
\item If $V^{j}_{n, t+i}$ is the approximation for $v_{t+i}$ in iteration $j$, then the next iterate $V^{j+1}_{n, t+i}$ is implicitly defined by Type I iteration (according to  Fair and Taylor's 
definition)
\[V^{j+1}_{n, t+i}= - B^{-1}G(u_{t+i}, V^{j+1}_{n, t+i}) + B^{-1}V^{j}_{n, t+i+1}.\]
\item Repeat Step 3 for $j =1,\dotsc, n$. These iterations are called Type II iterations.
\end{enumerate}
The first iteration of Type II, i.e. $j=1$, gives the approximation 
\[V^{1}_{n, t+i}=h_{1}(u_{t+i}), i=0,1,\dotsc, n,\]
where $h_{1}$ is the mapping defined in \eqref{eq:25}.
Therefore, the values $V^{1}_{n, t+i}$ equal the values of the mapping $h_1$ at the  points $u_{t+i}$. Completing $n$ iterations of Type II results in
\[V^{n}_{n, t}=h_{n}^{*}(u_{t}), 
\ldots,V^{n}_{n, t+i}=h_{n-i}^{*}(u_{t+i}),\dots, V^{n}_{n, t+n-1}=h_{1}^{*}(u_{t+n}), \]
where $h_{n-i}^{*}$, $i=0,1,\dots,n-1$, are the mappings that are defined by the following recurrent equations:
\begin{equation}
h_{k}^{*}(u)=-B^{-1}G(u,h_{k}^{*}(u)) +B^{-1}h_{k-1}^{*}(Au+F(u)),\label{eq:18__}
\end{equation}
where $k=1,\dots,n.$
This recurrent sequence of mappings corresponds to \eqref{eq:21} with the assumption that the variable $x_t$ is exogenous. Hence, Theorems~\ref{Thm1} and ~\ref{Thm2}   can be applied straightforwardly  for proving the convergence of the EP method. 
To the author's knowledge, until now there has been no proof that the EP algorithm converges to the true rational expectation solution to nonlinear models (see concluding remarks in \cite{r27}).  Therefore, the proposed approach can be considered as a rigorous proof that the EP algorithm converges to the true solution in a class of nonlinear models.

Moreover, the use of linearization at the steady state and spectral decomposition in the ASM approach gives us additional advantages  over the EP method in terms of the speed of convergence and size of the domain where the algorithms converge. Indeed, the EP method can hardly be considered
satisfactory in handling models with near-unit-root stable eigenvalues at the steady state (see, for example, \cite{r56}). This occurs because if the  Jacobian matrix at the steady state has  near-unit-root stable eigenvalues, then the Lipschitz constant of the corresponding  mapping exceeds one soon when the radius of neighborhood increases. The  Lipschitz constant greater than one does not allow for the convergence of the EP method in the domain by violating the assumption of the Contracting Mapping Theorem. 
Whereas in the ASM approach under the transformation of the system, the Jacobian  matrix at the steady state is a zero matrix, thus the radius of the domain, where the Lipschitz constant is less than one, is  likely larger  for the ASM method than for the EP one.

Taking into account the analogy with the EP method,  stochastic simulation for the ASM method can be done  in the same way as in \cite{r15}. Namely, under a given initial condition $(x_t,z_t)$ and the assumption that the disturbances  $\varepsilon_{t+i}$, $i \in \mathbb{N}$, are equal to zero in all future periods,  the deterministic system can be solved by 
the ASM method described in Section~\ref{theory}.
This solution also contains the next period initial condition for the variable $x_{t+1}$. 
The next period initial condition for exogenous state variable $z_{t+1}=\Lambda z_t+\varepsilon_{t+1}$ is obtained by drawing the stochastic disturbance $\varepsilon_{t+1}$. This provides the next period's control variable $y_{t+1}$. Iterating on this process yields a time-series realization of  stochastic simulation. As the EP method the AMS approach
neglects Jensen's inequality by setting future shocks to zero. 
Nevertheless, as shown in \cite{r40, r41}; and \cite{ r42}  the error induced by this approximation of future shocks may be quite small.


%
%
%


\section{Example. The Neoclassical growth model}\label{example}
\subsection{Implementation the solution method }
To illustrate how the presented method works we apply it to 
the simple deterministic
Neoclassical Growth Model of \cite{r2}. Consider the one-sector
growth model with inelastic labor supply. The representative agent maximizes the intertemporal utility function
\begin{equation}
\underset{\left\{ c_{t}\right\} }{\max }\sum _{t=0}^{\infty }\beta ^{t}\ln \left(
c_{t}\right)   \label{4.1}
\end{equation}
subject to%
\begin{equation}
c_{t}+k_{t}=k_{t}^{\alpha }.  \label{4.2}
\end{equation}
Using the resource constraint to substitute out consumption, we have the
following equilibrium condition:
\begin{equation}
\frac{1}{k_{t}^{\alpha }-k_{t+1}}=\beta \frac{\alpha }{\left(
k_{t+1}^{\alpha }-k_{t+2}\right) k_{t+1}^{1-\alpha }}.  \label{4.3}
\end{equation}
Inverting (\ref{4.3}) gives
\begin{equation*}
k^{\alpha}_{t}-k_{t+1}= \frac{1}{\beta\alpha}(k^{\alpha}_{t+1}-k_{t+2})k^{1-\alpha}_{t+1}.
\end{equation*}
Expressing $k_{t+2}$ as a function of $k_{t+1}$ and $k_t$ yields
\begin{equation}
k_{t+2} = \frac{(1+\alpha\beta)k_{t+1} -\alpha\beta k^{\alpha}_{t}}{k^{1-\alpha}_{t+1}}.\label{A5}
\end{equation}
Taking into account the steady state 
\begin{equation}
\bar{k} = (\beta\alpha)^{\frac{1}{1-\alpha}} ,\label{A6}
\end{equation}
we get 
\begin{equation}
{\hat{k}}_{t+2} = (1+ \alpha\beta)\left(\bar{k}+ {\hat{k}}_{t+1}\right)^{\alpha}-\frac{\alpha\beta\left(\bar{k}+ {\hat{k}}_{t}\right)^{\alpha}}                 
{\left(\bar{k}+ {\hat{k}}_{t+1}\right)^{1-\alpha}}  -  \bar{k}.\label{A7}
\end{equation}
Equation (\ref{A7}) can be rewritten as 
\begin{equation}
\begin{split}
{\hat{k}}_{t+2}& = (1+ \alpha\beta){\bar{k}}^{\alpha}\left(1+\frac{ {\hat{k}}_{t+1}}{\bar{k}}\right)^{\alpha} - \frac{\alpha\beta{\bar{k}}^{\alpha}\left(1+\frac{ {\hat{k}}_{t}}{\bar{k}}\right)^{\alpha}}                 
{{\bar{k}}^{1-\alpha}\left(1+\frac{ {\hat{k}}_{t+1}}{\bar{k}}\right)^{1-\alpha}}  -  \bar{k}\\
 &= f\left({\hat{k}}_{t+1},{\hat{k}}_{t}\right).\label{A8}
\end{split}
\end{equation}
It can easily  be checked that the derivatives of $f$ at the point $(0,0)$ are
\[f_{\hat{k}_{t}}^{'}(0,0)=-1/\beta\quad\mbox{and}\quad f_{\hat{k}_{t+1}}^{'}(0,0)=(1/\alpha\beta)+\alpha ,\]
therefore (\ref{A8}) takes the form
\begin{equation}
\begin{split}
&{\hat{k}}_{t+2} = \left(\frac{1}{\alpha\beta}+\alpha\right){\hat{k}}_{t+1}-\frac{1}{\beta}{\hat{k}}_{t}+\Bigg[(1+ \alpha\beta){\bar{k}}^{\alpha}\left(1+\frac{ {\hat{k}}_{t+1}}{\bar{k}}\right)^{\alpha}\\
& - \frac{\alpha\beta{\bar{k}}^{\alpha}\left(1+\frac{ {\hat{k}}_{t}}{\bar{k}}\right)^{\alpha}}                 
{{\bar{k}}^{1-\alpha}\left(1+\frac{ {\hat{k}}_{t+1}}{\bar{k}}\right)^{1-\alpha}}-  \bar{k}+\frac{1}{\beta}{\hat{k}}_{t}-\left(\frac{1}{\alpha\beta}+\alpha\right){\hat{k}}_{t+1}\Bigg].\label{A9}
\end{split}
\end{equation}
If we denote ${\hat{k}}_{t+1}$ by ${\hat{z}}_{t}$, then (\ref{A9}) can be rewritten as the following system of equations:
\begin{equation}
\begin{aligned}
&{\hat{k}}_{t+1} ={\hat{z}}_{t},\\
&{\hat{z}}_{t+1}= -\frac{1}{\beta}{\hat{k}}_{t}+\left(\frac{1}{\alpha\beta}+\alpha\right){\hat{z}}_{t}+N\left({\hat{k}}_{t},{\hat{z}}_{t}\right),\label{A10}
\end{aligned}
\end{equation}
where the nonlinear term is
\begin{equation*}
\begin{split}
N\left({\hat{k}}_{t},{\hat{z}}_{t}\right)&=(1+ \alpha\beta){\bar{k}}^{\alpha}\left(1+\frac{{\hat{z}}_{t}}{\bar{k}}\right)^{\alpha} - \frac{\alpha\beta{\bar{k}}^{\alpha}\left(1+\frac{ {\hat{k}}_{t}}{\bar{k}}\right)^{\alpha}}                 
{{\bar{k}}^{1-\alpha}\left(1+\frac{{\hat{z}}_{t}}{\bar{k}}\right)^{1-\alpha}}\\
& -  \bar{k}+\frac{1}{\beta}{\hat{k}}_{t}-\left(\frac{1}{\alpha\beta}+\alpha\right){\hat{z}}_{t}.
\end{split}
\end{equation*}
Rewriting (\ref{A10}) in the matrix form, gives
\begin{equation}
\begin{pmatrix}\hat{k}_{t+1}\\\hat{z}_{t+1}\end{pmatrix}=
K\begin{pmatrix}\hat{k}_{t}\\\hat{z}_{t}\end{pmatrix}+\begin{pmatrix}0\\N(\hat{k}_{t}, \hat z_{t})\end{pmatrix},\label{A11}
 \end{equation}
where the matrix $K$ has the form
 \[K = \left(\begin{array}[c]{ccc}%
0&1\\
-\frac{1}{\beta}&\frac{1}{\alpha\beta}+\alpha%
\end{array}
\right).\]
Next $K$ is transformed into Jordan canonical form $K=ZPZ^{-1}$ , where 
 \[P = \left(\begin{array}[c]{ccc}%
\alpha&0\\
0&\frac{1}{\alpha\beta}%
\end{array}
\right), 
 Z= \left(\begin{array}[c]{ccc}1&1\\
\alpha&\frac{1}{\alpha\beta}
\end{array}
\right), 
Z^{-1}= \frac{\alpha\beta}{1-\alpha^2\beta}\left(\begin{array}[c]{ccc}%
\frac{1}{\alpha\beta}&-1\\
-\alpha&1%
\end{array}
\right).\]
After introducing new variables  $ \left(\begin{array}[c]{ccc}%
u_t\\
v_t%
\end{array}
\right)=Z^{-1}\left(\begin{array}[c]{ccc}%
{\hat{k}}_{t}\\
{\hat{z}}_{t}%
\end{array}
\right)$ and premultiplying both side of (\ref{A11}) by $Z^{-1}$, we can rewrite  (\ref{A5}) as
\begin{equation*}\label{eq:A12}
\begin{aligned}
u_{t+1} = Au_{t} + F(u_{t}, v_{t})\\
v_{t+1} = Bv_{t} + G(u_{t}, v_{t}),
\end{aligned}
\end{equation*}
where $A=\alpha, B=1/(\alpha\beta), F(u_t,v_t)=Z^{12}N(Z_{11}u_t+Z_{12}v_t,Z_{21}u_t+Z_{22}v_t),$ and $G(u_t,v_t)=Z^{22}N(Z_{11}u_t+Z_{12}v_t,Z_{21}u_t+Z_{22}v_t)$, where $Z_{ij}$ and $Z^{ij}$ are the components of the matrices $Z$ and $Z^{-1}$, respectively. 

The mapping $G$ has the following  representation:
\begin{equation*}
\begin{split}
G(u_{t},v_{t})& = \frac{\alpha\beta}{1-\alpha^2\beta}
\Biggl[(1+ \alpha\beta)(\bar{k} +\alpha u_t+\frac{1}{\alpha\beta}v_t)^{\alpha} - \frac{\alpha\beta\left(\bar{k} +u_t+v_t\right)^{\alpha}}{(\bar{k} +\alpha u_t+\frac{1}{\alpha\beta}v_t)^{1-\alpha}}\\
& - \bar{k} - \left(\frac{1}{\alpha\beta}+\alpha\right)\left(\frac{1}{\alpha\beta}v_t+\alpha u_t\right)+\frac{1}{\beta}(u_t+v_t)\Biggr]
\end{split}
\end{equation*}
Then, the approximate policy function $h_1$ has the implicit form:
\begin{equation}
\begin{split}
h_1(u_{t})&= -\frac{\alpha\beta}{1-\alpha^2\beta}
\Biggl[(1+ \alpha\beta)(\bar{k} +\alpha u_t+\frac{1}{\alpha\beta}h_1(u_{t}))^{\alpha} \\
&- \frac{\alpha\beta\left(\bar{k} +u_t+h_1(u_{t})\right)^{\alpha}}{(\bar{k} +\alpha u_t+\frac{1}{\alpha\beta}h_1(u_{t}))^{1-\alpha}}- \bar{k} \\
&- \left(\frac{1}{\alpha\beta}+\alpha\right)\left(\frac{1}{\alpha\beta}h_1(u_{t})+\alpha u_t\right)+\frac{1}{\beta}(u_t+h_1(u_{t}))\Biggr].\label{A13}
\end{split}
\end{equation}

The first iteration, $h_{1,1}$, of the contracting mapping is obtained by substituting zeros for $h_{1}(u_t)$ in the right hand side of (\ref{A13})
\begin{equation*}
\begin{split}
h_{1,1}(u_{t}) = -\frac{\alpha\beta}{1-\alpha^2\beta}
\Biggl[(1+ \alpha\beta)(\bar{k} +\alpha u_t)^{\alpha} - \frac{\alpha\beta\left(\bar{k} +u_t\right)^{\alpha}}{(\bar{k} +\alpha u_t)^{1-\alpha}}- \bar{k} - \alpha^2 u_t\Biggr].\label{14}
\end{split}
\end{equation*}
The function $h_1$ can be found by further iteration of the right hand side of \eqref{A13}. 
In the same way, the functions $h_2$ and $h_3$ satisfy the following equations:
\begin{equation*}
h_{2}(u)=-B^{-1}G(u,h_{2}(u)) +B^{-1}h_{1}(Au+F(u,h_{2}(u))).
\end{equation*}
\begin{equation*}
h_{3}(u)=-B^{-1}G(u,h_{3}(u)) +B^{-1}h_{2}(Au+F(u,h_{3}(u))).
\end{equation*}
Having mappings $h_i$, $i=1$, $2$
, or $h_{1,1}$, to return to the original variables we must perform the transformation: 
\begin{equation*}
\begin{pmatrix}k_{t}\\k_{t+1}\end{pmatrix}=
Z\begin{pmatrix}{u}_{t}\\h^{*}(u_{t})\end{pmatrix}+\begin{pmatrix}\bar{k}\\\bar{k}\end{pmatrix},
\end{equation*}
where $h^{*}=h_i$ or $h_{1,1}$.
Finally, the policy function $k_{t+1}=\tilde{h}(k_t)$ has the following parametric representation:
\begin{align*}
\begin{cases}
k_t&=u_t+h^{*}(u_t)+\bar{k}\\
\tilde{h}(k_t)&=\alpha u_t+\frac{1}{\alpha\beta}h^{*}(u_t)+\bar{k}.
\end{cases}
\end{align*}

\subsection{Global approximation of the ASM method}
To give an impression of the accuracy  of the approximations for the ASM method on a global domain we construct  the functions $\tilde{h}_{1}$ and $\tilde{h}_{2}$ 
from Section~\ref{theory}
and compare them with  the Taylor series expansion of the 1st, 2nd, 5th and 16th-orders, which correspond to the solutions of perturbation methods by \cite{ r14} in the case of zero shock volatilities. This model has a closed-form solution for the policy function 
given
by 
\begin{equation}
k_{t+1}=\tilde{h}(k_{t})=\alpha \beta k_{t}^{\alpha }.  \label{4.4}
\end{equation}
We calculate approximations in the level (rather than in the logarithm) of the state
variable, otherwise the problem becomes trivially linear. 
\begin{figure} 
\includegraphics [scale=.550]  {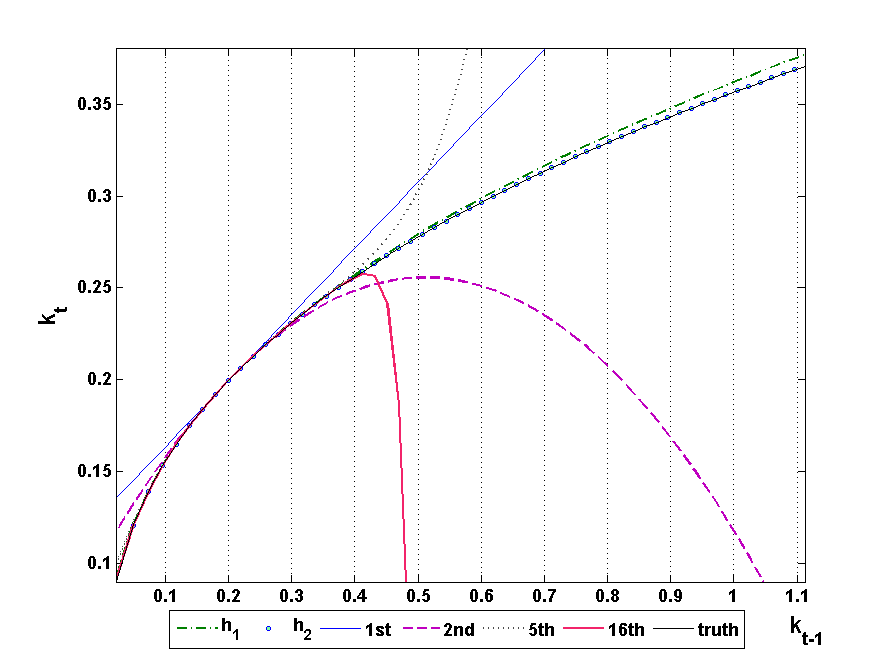}
\caption[]{ Policy functions for different approximations.}
\label{Fig1}
\end{figure}
The parameter
values take on standard values, namely $\alpha =0.36$ and $\beta=0.99$. Then for our calibration the steady state of capital is  $\bar{k}=\left(
\alpha \beta \right) ^{1/(1-\alpha )}=0.20$.  It is not hard to see that the
Taylor series expansion of the true solution (\ref{4.4}) converges in the interval $[0,2\bar{k}]$.

Figure~\ref{Fig1} shows different solutions of the capital policy
function  on the global interval $[0, 5\bar{k}]$. 
The Taylor
series approximation perform well in the interval $[0,0.40]$, i.e. within
the domain of the convergence of the Taylor series expansion; however, outside the interval they explode.  The
functions $\tilde{h}_{2}$ 
is essentially indistinguishable with the
true solution for all $k_{t}$, thus providing  the perfect global
approximation. Even within the domain of convergence of the Taylor series the accuracy of $\tilde{h}_{2}$ is as good as for the 16th-order of the Taylor series expansion. The function $\tilde{h}_{1}$ also provides a very close fit for the whole interval.

%
%
%

\section{Conclusion}\label{conclusion}
This study has used concepts and technique originated from dynamical system theory to construct approximate solutions to general equilibrium models on nonlocal domains and proved the convergence of these approximations to the true solution. As a result, a new proof of the Stable Manifold Theorem is obtained. 
%
The proposed method allows estimating the a priory and a posteriori approximations errors since it involves the Contraction Mapping Theorem. As a by-product, the approach can be treated as a rigorous proof of convergence of the EP algorithm.
The method is illustrated by applying to the neoclassical growth model of \cite{r2}. The results shows that just a first few approximations of the proposed method  are very accurate globally, at the same time they have at least the same accuracy locally as the solutions obtained by the higher-order perturbation method.   


\begin{thebibliography}{}

%
%


\bibitem[\protect\citeauthoryear{Adjemian and  \mbox{Juillard}}{2011}]{r40}
\textsc{Adjemian, S.}, and \textsc{ Juillard, M.} (2011).
Accuracy of the extended path simulation method
in a new keynesian model with zero lower bound on the nominal interest rate." Mimeo, Universit\'{e} du Maine.

%

\bibitem[Blanchard and Kahn (1980)]{r25}
\textsc{Blanchard, O.J.}, and \textsc{Kahn, C.M.}  (1980).
The Solution of Linear Difference Models Under Rational Expectations, \textit{ Econometrica}
\textbf{48} 1305--1311.

\bibitem[\protect\citeauthoryear{Brock and Mirman}{1972}]{r2}
\textsc{Brock, W.A.} and \textsc{Mirman, L.} (1972).
Optimal economic growth and uncertainty. the discounted case,'' \textit{Journal of Economic Theory}
\textbf{4} 479--513.

\bibitem[\protect\citeauthoryear{Collard and Juillard}{2001}]{r3}
\textsc{Collard, F.}, and \textsc{Juillard, M. } (2001).
Accuracy of stochastic perturbation methods: the case of asset pricing models, \textit{Journal of Economic Dynamics and Control }
\textbf{25} 979--999.
%

\bibitem[\protect\citeauthoryear{Fair and Taylor}{1983}]{r15}
\textsc{Fair, R.}, and \textsc{Taylor, J.B. } (1983).
Solution and maximum likelihood estimation of dynamic rational expectation models, \textit{Econometrica}
\textbf{51} 1169--1185.


\bibitem[\protect\citeauthoryear{Gagnon}{1990}]{r41}
\textsc{Gagnon, J.E.} (1990).
Solving Stochastic Equilibrium Models with the Extended Path Method,''\textit{Journal of Business and Economic Statistics}
\textbf{8} 35--36.

\bibitem[\protect\citeauthoryear{Gagnon and Taylor}{1990}]{r27}
\textsc{Gagnon, J.E.}, and \textsc{Taylor, J.B. } (1990).
Solving the stochastic growth model by deterministic
extended path,''\textit{Economic Modelling}
\textbf{7} 251--257.




\bibitem[\protect\citeauthoryear{Golub and Van Loan }{1996}]{r26}
\textsc{Golub, G.H.}, and {Van Loan, C.F.} (1996). \textit{Matrix Computations}, 3rd ed.
Johns Hopkins University Press, Baltimore.
%

\bibitem[\protect\citeauthoryear{Hartmann}{1982}]{r6}
\textsc{Hartmann, P.} (1982). \textit{Ordinary Differential Equations}, 2nd ed.
Wiley, New York.

\bibitem[\protect\citeauthoryear{Jin and  Judd }{2002}]{r7}
\textsc{Jin, H.}, and \textsc{ Judd, K. L. } (2002).
Perturbation methods for general dynamic stochastic models, Discussion Paper, Hoover Institution, Stanford.

\bibitem[\protect\citeauthoryear{Judd}{1998}]{r8}
\textsc{Judd, K. L.} (1998). \textit{Numerical Methods in Economics}, 
The MIT Press, Cambridge.


\bibitem[\protect\citeauthoryear{Katok, and Hasselblatt}{1995}]{r9}
\textsc{Katok A.} and \textsc{Hasselblatt B.}(1995). \textit{Introduction to the modern theory of dynamical systems.}, 
Cambridge Univ. Press, Cambridge.


\bibitem[\protect\citeauthoryear{King and Watson}{2002}]{r11}
\textsc{King, R.G.} and \textsc{ Watson, M.} (2002).
System reduction and solution algorithms for singular linear difference systems under rational expectations, \textit{Computational Economics}
\textbf{20}57--86.

\bibitem[\protect\citeauthoryear{Laffargue}{1990}]{r56}
\textsc{Laffargue, J.- P.} (1990).
R\'esolution d'un mod\`ele macro\'economique avec anticipations
rationnelles, \textit{Annales d'Economie et Statistique}
\textbf{17}  97--119.

\bibitem[\protect\citeauthoryear{Love}{2010}]{r42}
\textsc{Love, D.R.F.} (2010).
Revisiting deterministic extended-path: a simple and accurate solution
method for macroeconomic models, \textit{Int. J. Computational Economics and
Econometrics}, 
\textbf{1}309--316.

\bibitem[\protect\citeauthoryear{Lipton et al.}{1982}]{r13}
\textsc{Lipton D.}, \textsc{Poterba J.}, \textsc{Sachs J.} and \textsc{Summers L.} (1982).
Multiple Shooting in Rational Expectations Models, \textit{Econometrica}
\textbf{50} 1329--1333.

\bibitem[\protect\citeauthoryear{Nitecki}{1971}]{r31}
\textsc{Nitecki, Z.} (1971). \textit{Differentiable Dynamics}, 
The MIT Press, Cambridge.

\bibitem[\protect\citeauthoryear{Potzsche}{2010}]{r28}
\textsc{Potzsche C.} (2010). \textit{Geometric Theory of Discrete Nonautonomous Dynamical Systems}, 
Springer-Verlag, Berlin-Heidelberg-New-York-Tokyo. 

\bibitem[\protect\citeauthoryear{Schmitt-Groh\'{e} and Uribe}{2004}]{r14}
\textsc{Schmitt-Groh\'{e}, S.}, and \textsc{Uribe, M.} (2004).
Solving dynamic general equilibrium models using as second-order approximation to the policy function, \textit{Journal of Economic Dynamics and Control}
\textbf{28} 755--775.

\bibitem[\protect\citeauthoryear{Smale}{1967}]{r30}
\textsc{Smale, S.} (1967).
Differentiable dynamical systems,  \textit{Bulletin of the American Mathematical Society}
\textbf{73} 747--817.

\bibitem[\protect\citeauthoryear{Stuart}{1994}]{r44}
\textsc{Stuart, A.M.} (1990).
Numerical Analysis of Dynamical Systems,  \textit{Acta Numerica}
\textbf{3} 467--572.


\bibitem[\protect\citeauthoryear{Zeidler}{1986}]{r22} 
\textsc{Zeidler E.} (1986). \textit{Nonlinear functional analysis and its applications,  Vol. I.},
Springer-Verlag, Berlin-Heidelberg-New-York-Tokyo. 

\end{thebibliography}
\end{document}